%% file: CBM.tex
\documentclass[12pt]{article}
\usepackage{epsfig}

\input econfmacros-csqcd3.tex
\def\Title#1{\begin{center} {\Large {\bf #1} } \end{center}}

\begin{document}

\Title{Compressed Baryonic Matter of Astrophysics}

\bigskip\bigskip


\begin{raggedright}

{\it Yanjun Guo {\rm and} Renxin Xu\\
School of physics\\
Peking University\\
Beijing 100871\\
P. R. China\\
{\tt Email: r.x.xu@pku.edu.cn}}
\bigskip\bigskip
\end{raggedright}

\abstract{
Baryonic matter in the core of a massive and evolved star is compressed significantly to  form a supra-nuclear object, and compressed baryonic matter (CBM) is then produced after supernova.
The state of cold matter at a few nuclear density is pedagogically
reviewed, with significant attention paid to a possible
quark-cluster state conjectured from an astrophysical point of view.
}

\section{An introduction to CBM}

It is well known from astrophysical observations that our universe
is dominated by dark matter and dark energy, with a fraction of
$\sim 23$\% and $\sim 73$\%, respectively.
Nevertheless, the most familiar composition we know best is the 4\%
atomic part---the baryons and leptons!
In the standard model of particle physics, all of matter are composed
of fundamental Fermions, 6 flavors of quarks and 6 flavors of leptons,
between which elementary interactions are mediated by gauge bosons.
Particles made of quarks are called baryon, and the baryonic number of
a quark is 1/3.
With Higgs boson, the origin of mass, discovered by LHC experiments
at 5$\sigma$ confidence level in July 2012, all 62 particles within
the frame of standard model have been evidenced.

The most familiar and stable baryon are nucleon (proton and
neutron), which is made of up and down quarks, the lightest two
flavors of quarks easiest to be excited.
At low energy scale, protons and neutrons are strongly coupled by
residual forces of color interaction between quarks and form various
kinds of atom nuclei.
Nucleus with positive charge attracts electrons and then constitute
atom combined by electromagnetic force, which is the building-block
of ordinary substances.
Although the nuclei contribute $> 99.9$\% of the atomic mass,
its length scale is only $\sim 10^{-13}$ cm = 1 fm, while that of
atom is order of \AA\, $= 10^{-8}$ cm; this means that there is
plenty of empty space between atoms.
%
%
Possible matter at the highest density, limited by the sizes of
electrons and nuclei, was speculated by Fowler in 1926
~\cite{Fowler1926}.
If the space is squeezed out, one would then obtain the so-called
compressed baryonic matter (CBM), which could be of extremely high
density, even larger than the nuclear density, $\rho_{\rm nucl}$.
Although it is almost impossible yet to squeeze the space out of
normal matter in terrestrial laboratory, strong gravity inside the
core of an evolved massive star could do this: the fascination of
astrophysics!
CBM could thus be in the heaven, and supernova could make CBM
from atoms~\cite{BZ1934}.

The outline of this paper is as follows.
In \S2, we will start with what we know about CBM, and what we
do not know about CBM will be considered in \S3.
A possible state of CBM, quark-cluster, is discussed in \S4.
Then, in \S5, observational hints for the nature of CBM are summarized.
Finally, we make conclusions in \S6.

\section{What we do know about CBM}

Let us put aside various speculations conceivable or not, and just think
about what we are sure of CBM first.

For a pulsar-like compact star with a mass of $\sim M_\odot$, the
baryon number of CBM inside can be roughly estimated to be $\sim
M_\odot/m_{\rm p}\sim 10^{57}$ ($m_{\rm p}$ is the proton mass),
which is so large that the medium effect would be significant.
If nuclei are considered as gravity-free, the density of CBM would
be larger than $\rho_{\rm nucl}$ as a consequence of gravitational
compression.
CBM is so dense that even a nugget of CBM not larger than a rubber
could be as heavy as all of the world's population.

We can also estimate the energy scale of quarks in CBM via two
simple ways.
(i) In case of free quarks, the Fermi momentum is
$p_F=(3\pi^2)^{1/3}\hbar n^{1/3}$, and a calculation of Fermi energy
gives,
\begin{equation}\label{}
  E_F^{NR} \approx \frac{\hbar^2}{2m_q} (3\pi^2)^{2/3} \cdot n^{2/3} =380 \textrm{MeV}
\end{equation}
if quarks are considered moving non-relativistically, or
\begin{equation}\label{}
  E_F^{ER} \approx \hbar c (3\pi^2)^{1/3} \cdot n^{1/3} = 480 \textrm{MeV}
\end{equation}
if quarks are considered moving extremely relativistically, where
$n\simeq (3\times 0.16 = 0.48)$ fm$^{-3}$ is the number density of
each flavor of quark and the dressed quark mass $m_q\simeq 300$ MeV.
(ii) In case of localized quarks, one could also estimate the
zero-point momentum by Heisenberg's uncertainty relation, $p_0
\approx \hbar n^{1/3} \sim p_F$.
Therefore, we have the energy scale $E_{\rm scale}\sim 400$ MeV by
either Heisenberg's relation or Fermi energy.

Such a energy scale would have two implications.

(1) CBM should be {\em strange}.
As $E_{\rm scale}$ would be much larger than the mass difference
between strange and up/down quarks, $\Delta m \sim 100 \textrm{
MeV}$, strange quark could be easily provoked and strangeness may
play an important role in determining the nature of CBM, while
heavier quarks ($c,t,b$) are not likely to be excited.

(2) Non-perturbative QCD effects are significant for CBM because
$E_{\rm scale}<1$ GeV. The state of realistic CBM should be far from
the region where the asymptotic freedom approximation could apply.
It is worth noting that the the strong coupling between quarks might
render quarks grouped, although this hypothetical quark-cluster~\cite{Xu03}
in CBM has not been confirmed due to the lack of both theoretical and
experimental evidence.
For a quark-cluster with length scale $l$, from Heisenberg's
relation, the kinetic energy would be of $\sim p^2/m_q \sim
\hbar^2/(m_q l^2)$, which has to be comparable to the color
interaction energy of $E\sim \alpha_s \hbar c/l$ in order to have a
bound state, where $\alpha_s$ is the coupling constant of strong
interaction.
One then finds if quarks are dressed,
\begin{equation}\label{}
 l \sim \frac{1}{\alpha_s} \frac{\hbar c}{m_q c^2} \simeq \frac{1}{\alpha_s} \textrm{ fm, }
 E \sim \alpha_s^2 m_q c^2 \simeq 300 \alpha_s^2 \textrm{ MeV.}
\end{equation}
It is evident that the interaction energy, $E$, would be approaching
and even larger than the Fermi energy ($\sim$ 400 MeV) if
$\alpha_s>1$ for CBM at a few $\rho_{\rm nucl}$, which means that
quarks in CBM might not behave like Fermi gas.

In summary, for CBM manifesting as pulsar-like compact stars, we
know that there are $\sim 10^{57}$ quarks (with strangeness?), and
non-perturbative color interaction should play an important role in
determining the equation of states (EoS).
But what can we understand more about CBM?

\section{What we do not know about CBM}
There are many things that we do not know about CBM, as CBM is
hard to be created in terrestrial laboratory, and a direct calculation
from first principle is difficult due to non-perturbative effects of
low-energy QCD.

To begin with, a challenging problem is whether the quarks are
confined or de-confined in CBM, according to which different
scenarios are suggested for pulsar-like compact stars.
%
%
In hadron star model~\cite{Landau1932, Hewish1968, Gold1968},
quarks are confined in hadrons such as neutron/proton and hyperon,
while quark matter would exist in a core of hybrid/mixed star~\cite{IK1969}
whose central density could be high enough to make quarks de-confined.
%
A quark star is a condensed object dominated by free quarks
~\cite{Itoh1970, Witten1984, Alcock1986, Haensel1986}.
%
%
Strictly speaking, a quark-cluster star, however, is neither a
hadron star nor a quark star, in which strong coupling cause
individual quarks grouped in clusters.
Among these scenarios, hadron star and hybrid star are gravity-bound,
which are covered by crusts with nuclei and electrons,
while quark star and quark-cluster star are self-bound on surface.
This surface difference may have profound implications for
observations.

There exists another essential problem: would 3-flavor symmetry
be restored in CBM?
In an ordinary nucleus, a symmetry is kept between proton \{uud\}
and neutron \{udd\}, which is essentially equivalent to a 2-flavor
symmetry.
As an up quark carries $+2/3$ charge and a down quark just $-1/3$
charge, to keep electric neutrality, electrons as many as u/d quarks
would participate in matter with 2-flavor symmetry.

In ordinary case, electrons are outside the nucleus and their energy
$E_e$ is far less than 1 MeV, so atoms could be stable with 2-flavor
symmetry.
Nevertheless, things are different in case of CBM, for that
electrons are inside the gigantic nucleus and the kinematic energy
of electron would be $E_e \sim 100$MeV.
Such a high energy might intensify the interaction $e+p \to
n+\nu_e$, thus $E_e$ decreases but, unfortunately, the nuclear
symmetry energy $E_{\rm sym}$ increases.
This embarrassed situation does not exist if strange quark is invoked
in CBM (gigantic nucleus); the number of electrons there could be
only $10^{-5}$ less than that of u/d quarks as $s$ quark is heavier.

If 3-flavor symmetry is restored, the number of electrons in CBM
would be much less, which makes $E_e \sim 10$ MeV, and the
gigantic nucleus would be stable.
Certainly the argument above is not suitable for ordinary nucleus,
as the surface energy would increase with decreasing radius, and
different from gigantic nucleus, electrons are outside the ordinary
one, which causes a much smaller kinetic energy not sufficient for
$s$ quark to be excited.
%

From the theoretical arguments above, there is a possibility that
CBM could be strange quark-cluster matter, which distinguishes
from neutron star matter and quark matter.
%

\section{What if CBM is made of quark-clusters?}

If CBM is composed of strange quark-cluster matter, there would be mainly
three consequences.

First, CBM would have a stiff EoS, because quark-cluster should be
non-relativistic particle for its large mass, and because there
could be strong short-distance repulsion between quark-clusters.
As we all know, the relation between energy and momentum is
$E=(c^2p^2+m^2c^4)^{1/2}$, which can be approximated as $E=p^2/2m$
in non-relativistic (NR) case, while $E=pc$ for extra-relativistic
(ER) case.
From relations above we can get the EoS $P \sim \rho^\gamma$, and
$\gamma$ would be larger in NR case, which means a stiffer EoS.
As for interaction between quark-cluster matter, recently,
H-dibaryon has been
found in lattice QCD simulations by two independent
groups, with binding energy of about 10 to 40 MeV~\cite{Beane10, Inoue10}.

Second, different from traditional neutron stars, quark-cluster star
would be self-bound by residual interaction between clusters, which
could be a crucial difference providing observational manifestations
to distinguish the two models.

Last but not least, CBM could be in a global solid state if the
kinetic energy $kT$ is less than interaction energy between
quark-clusters, where $k$ is the Boltzmann constant and $T$ is the
temperature.
Solid quark-cluster star could possess more free energy reserved as
elastic and gravitational ones, which might be alternative energy
sources for the bursts and even giant flares in anomalous X-ray
pulsars (AXP) and soft gamma-ray repeaters (SGR).

Except these qualitative features analyzed, how shall we model
quark-cluster star?
Certainly, it is extremely difficult to calculate from first
principles due to non-perturbative effects, but phenomenologically,
there may be some feasible ways to probe the properties of
quark-cluster matter.

Motivated by recent lattice QCD simulations, a possible kind of
realistic quark-cluster, H-cluster, is considered~\cite{H-star}.
Assuming that interaction between H-clusters is mediated by
$\sigma$-$\omega$ mesons, we can derive EoS as well as the
mass-radius relations of H-cluster stars in different cases of the
in-medium stiffening effect and surface density.
Under a wide range of parameter-space, the maximum mass of H-cluster
stars can be well above 2$M_\odot$, and the calculated mass-radius
relations are consistent with both observations of the massive
2$M_\odot$ pulsar PSR J1614-2230~\cite{Demorest10} and the rapid
burster MXB 1730-335~\cite{Sala12}.

We have also studied the properties of quark-cluster matter by a
corresponding-state approach~\cite{glx12}.
Considering a group of substances described by potential in this
form: $\varphi=\varepsilon f(r/\sigma)$, we can express macroscopic
quantities, such as pressure $P$, volume $V$ and temperature $T$, in
dimensionless terms,
\begin{eqnarray}
  P^* &=& P\sigma^3/\varepsilon \label{eq-p}, \\
  V^* &=& V/(N\sigma^3) \label{eq-v}, \\
  T^* &=& kT/\varepsilon.
\end{eqnarray}
Through constructing another dimensionless parameter
$\Lambda^*=h/(\sigma\sqrt{m\varepsilon})$, corresponding to the de
Broglie wavelength, to measure the importance of quantum effects,
it can be proved that the reduced EoS expressed in these dimensionless
quantities, $P^*=f(V^*,T^*,\Lambda^*)$, is a universal relation, as
formulated by the law of corresponding states~\cite{de Boer1948}.
For inert gases described by Lennard-Jones potential,
$\varphi=\varepsilon
\{\frac{4}{(r/\sigma)^{12}}-\frac{4}{(r/\sigma)^6}\}$, the universal
EoS could be acquired by their experimental data.
If quark-cluster matter could be analogized to inert gases, and the
corresponding $\varepsilon$ and $\sigma$ could be determined, we can
get the EoS of quark-cluster matter by employing the corresponding-state
approach.
According to the derived mass-radius relation, the maximum mass could
also be well above 2$M_\odot$ under reasonable parameters.

\section{What can observations teach us?}

The realistic state of CBM is very difficult to directly calculate
from first principles, nonetheless, pulsar-like compact stars are
excellent astrophysical laboratory, observations of which could give
us useful hints for the nature of CBM.

For example, radio observations of PSR J1614-2230, a binary
millisecond pulsar with a strong Shapiro delay signature, imply that
the pulsar mass is 1.97$\pm$0.04 $M_\odot$~\cite{Demorest10}, which
indicates a stiff EoS for CBM.
It is conventionally thought that the state of dense matter softens and thus
cannot result in high maximum mass if pulsars are quark stars, and that the
discovery of 2$M_\odot$ pulsar may make pulsars unlikely to be quark stars.
However, as shown by qualitative analysis and empirical calculations in
\S4, quark-cluster star could not be ruled out by PSR J1614-2230, and the
observations of pulsars with higher mass, e.g. $>3M_\odot$ , would even
be a support to our quark-cluster star model, and give further constraints to
the parameters.

In measurements of black hole mass distribution, a significant mass
gap between the maximum pulsar mass detected (2$M_\odot$) and
the low end of the black hole mass distribution ($\sim 5M_\odot$)
has been identified previously~\cite{Ozel12}, which may suggest that
the maximum mass of pulsar would be in the middle range of
$2M_\odot-5M_\odot$.
However, after revising systematic errors in the mass measurements,
GRO J0422+32 and 4U 1543-47 may have small black hole masses (below
$\sim (4-5)M_\odot$), which may doubt the mass gap identified in the
previous work~\cite{Kreidberg12}.

Although both neutron star and quark-cluster star could account for
the discovery of high mass pulsars, many observation phenomena might
imply a self bound surface for CBM.
Drifting subpulses phenomena in radio pulsars suggest the existence of
Ruderman-Sutherland-like gap-sparking and thus strong binding of particles
on pulsar polar caps to form vacuum gap, but the calculated binding energy
in neutron star models could not be so high unless the magnetic field is
extremely strong.
This problem could be naturally solved in quark-cluster star
scenario due to the strong self bound nature on
surface~\cite{Xu1999, Qiao04}.

In addition, many theoretical calculations predict the existence of atomic
features in the thermal X-ray emission of neutron star atmospheres, which
should be detectable by Chandra and XMM-Newton.
However, none of the expected spectral features has been detected with
certainty up to now, and such non-atomic thermal spectra of dead pulsars
also hint that there might not exist the atmospheres speculated in neutron
star models.
Though conventional neutron star models cannot be ruled out by only
non-atomic thermal spectra since modified atmospheres with very
strong surface magnetic fields~\cite{HL03, Turolla04} might
reproduce a featureless spectrum too, a natural suggestion to
understand the general observation is that pulsars are actually
quark-cluster star without atoms on the surface~\cite{Xu02}.

Additionally, the bare and chromatic confined surface of
quark-cluster star could overcome the baryon contamination problem
and create a clean fireball for $\gamma$-ray burst and supernova.
The strong surface binding would result in extremely energetic
exploding because the photon/lepton luminosity of a quark-cluster
surface is not limited by the Eddington limit, and supernova and
$\gamma$-ray bursts could then be
photon/lepton-driven~\cite{Ouyed05, Paczynski05, Chen07}.

In order to explain observations, one needs either neutron stars
with super strong magnetic fields (i.e., magnetars, $>10^{14}$ G) or
quark-cluster stars with self-bound surfaces and normal fields
($\sim 10^{12}$ G).
How shall we distinguish the two models?
As neutron star is gravity-bound while quark-cluster star self-bound
by chromatic interaction, there is atmosphere on the surface of
neutron star while not on quark-cluster star, and thus a larger
temperature gradient in the former case.
The neutron star atmosphere would be ionic polarized in strong
magnetic field, so its thermal radiation would be linearly polarized
($\sim$ 10\%), while the polarization degree of quark-cluster star
could be almost zero~\cite{LXF13}.
That means neutron star and quark-cluster star can be clearly
distinguished by measuring linear X-ray polarization of dead pulsars
with pure thermal radiation.

Except for hints from the surface, there are some observations implying
global properties.
As addressed before, quark-cluster stars would be in a global solid
state like ``cooked eggs'', while for normal neutron stars, only
crust is solid like ``raw eggs''.
Rigid body would precess naturally when spinning, either freely or
by torque, and the observation of possible precession or even free
precession of B1821-11~\cite{Stairs2000} and others could suggest a
global solid structure for pulsar-like stars.

Star-quake is a peculiar action of solid compact stars, during which huge
free energy, such as gravitational and elastic energy, would be released.
Assuming the two kinds of energies are of same order, for a pulsar
with mass $M$ and radius $R$, the stored gravitational energy is
$E_{\rm stored} \simeq GM^2/R \sim 10^{53}$ erg if $M\sim M_\odot$
and $R\sim 10$ km, so energy released would be as high as $\sim
10^{53} \Delta R/R$ erg when the radius changes from $R$ to ($R-\Delta
R$).
Compared with magnetars powered by magnetic energy, quake-induced
energy in solid quark-cluster stars may also power the bursts,
flares and even superflares of AXPs and SGRs~\cite{Xu06}.
The question whether there is strong magnetic field in pulsar-like
compact stars is encountered again, which could be answered by
observations of linear X-ray polarization.

Another observational hint for the nature of CBM comes from low-mass
compact stars, including km-radius compact stars and planet-mass
compact objects.
As we know, neutron stars are gravitationally bound, while quark-cluster
stars are bound not only by gravity but also by additional strong interaction.
This fact results in an important astrophysical consequence that
quark-cluster star can be of very low mass with small radii, while
neutron stars cannot.
Thermal radiation components from some pulsar-like stars are
detected, and the radii are usually much smaller than 10 km in
blackbody models, which suggests the existence of km-radius compact
stars~\cite{Pavlov04}.
Besides, the compact companion of PSR J1719-1438 with a Jupiter-like
mass is suggested to be an exotic quark object rather than a light
helium or carbon white dwarf~\cite{Horvath12}.

X-ray bursts on stellar surface are believed to be evidence for
curst, which could be well explained by elaborate modeling in
neutron star model.
Can it be reproduced in quark-cluster star model?
The key to the mechanism of X-ray bursts is to have unstable energy
release during accretion, which could be implemented in quark-cluster
star by either thermal nuclear flash on crust formed above quark-cluster
star or star-quake-induced burst.
Other phenomena such as cooling, glitch and braking could also
provide hints for the nature of CBM~\cite{Blaschke12, Negreiros12,
Glendenning1997, Anderson12}.

Various observations hint a new answer for the nature of CBM: a
solid state of quark-cluster matter.
In the future, further observations realized by more advanced telescopes
would teach us more.
In radio band, Chinese five hundred meter aperture spherical
telescope (FAST~\cite{fast11}), the biggest single-dish radio
telescope to be built, is capable to measure the mass and even the
inertial of momentum of radio pulsars, and possibly find
sub-millisecond radio pulsars.
In X-ray band, Chinese lightweight asymmetry and magnetism probe
(LAMP) is also designed to detect X-ray polarization, which may shed
light on the nature of astrophysical CBM.

\section{Conclusions}

Why should one study CBM of astrophysics?
Frank Wilczek's answer is recommended~\cite{Wilczek11}: ``because
it's there'' (a variation on the one George Mallory gave) and
because it is a mathematically well-defined domain (to understand
Yang-Mills theory).
It is challenging for both physicist and astrophysicist to solve the
problem.

The history of astrophysical CBM study can date back to about eighty
years ago, when Landau proposed the idea of gigantic nucleus.
The discovery of pulsar in 1967 is a breakthrough in the study, and
Landau's speculation gradually developed to the standard neutron
star model today.
It is also conjectured, from an astrophysical point of view, that
CBM would actually be quark-cluster matter, which could be necessary
to understand different manifestations of pulsar-like compact stars.
Besides this, we are expecting key observations to test the models.

\bigskip
We would like to thank many contributions and useful discussions of
members at the pulsar group of PKU.
This work is supported by the National Basic Research Programme of
China (Grant Nos. 2012CB821800, 2009CB824800) and the National
Natural Science Foundation of China (Grant Nos. 11225314, 10935001).



\end{document}

%% file: econfmacros-csqcd3.tex
\def\beq{\begin{equation}}
\def\eeq#1{\label{#1}\end{equation}}
\def\eeqn{\end{equation}}

\def\beqa{\begin{eqnarray}}
\def\eeqa#1{\label{#1}\end{eqnarray}}
\def\eeqan{\end{eqnarray}}

\let\bar=\overbar

\def\Dslash{\not{\hbox{\kern-4pt $D$}}}
\def\dslash{\not{\hbox{\kern-2pt $\del$}}}

\def\msb{{\bar{\ssstyle M \kern -1pt S}}}

\usepackage{fancyhdr,graphicx}